# STANDARD MODEL HIGGS SEARCH IN THE FOUR JET CHANNEL AT LEP

G.J.DAVIES

*Imperial College of Science, Technology and Medicine, London*
On behalf of the LEP Collaborations

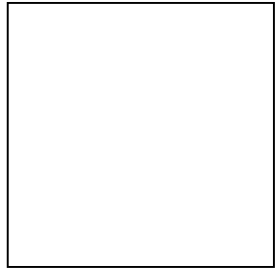

The Standard Model Higgs boson was searched for at LEP, by the four collaborations in the year 2000, at centre-of-mass energies from 200-209GeV. An excess of events, potentially compatible with the production of a Higgs boson of mass ~115GeV was observed, largely in the four-jet channel. The search in this channel is discussed in this letter.

## 1 Introduction

Results from the individual experiments are based upon their respective publications, produced soon after data taking ceased [1,2,3,4]. Combined results are based upon the combination of the Nov 3$^{rd}$ LEPC [5].

In the year 2000 LEP again performed excellently, delivering an integrated luminosity of ~870pb$^{-1}$ to the four experiments at centre-of-mass energies from 200-209GeV. As discussed in detail in these proceedings[6], Higgs production at LEP is predominately via Higgsstrahlung, with the dominant decay channel being $e^+e^- \to HZ \to b\bar{b}\,q\bar{q}$ (~55%). For a 114GeV Higgs one would expect ~3.5 four-jet events to be seen per experiment. Indeed the significance of the four-jet channel is the same as the other channels combined – as illustrated in figure 1.

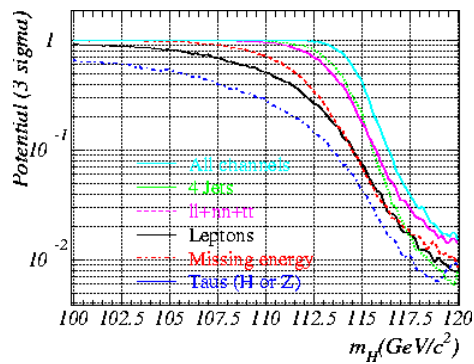

Figure 1 (above) 3σ search potential for the four different final states.

Issues of critical importance to the four-jet channel are b-tagging and jet pairing/ mass reconstruction, and are discussed in more detail below.

*1.1 B-tagging*

The four-jet channel uses the same b-tag as the other decay channels. Information on the b-hadron lifetime, mass and semi-leptonic decays are combined into a neural net. The neural nets are calibrated with the data taken each year at the Z peak and then checked at higher energies. High energy samples enriched in b-quarks can be obtained via the process $e^+e^- \rightarrow Z\gamma$ ; whilst samples enriched in non b-quarks can be obtained via the process $e^+e^- \rightarrow WW \rightarrow l\nu qq$. The agreement between data and Monte Carlo is shown in figure 2 for such samples, taken by DELPHI. All experiments perform these checks.

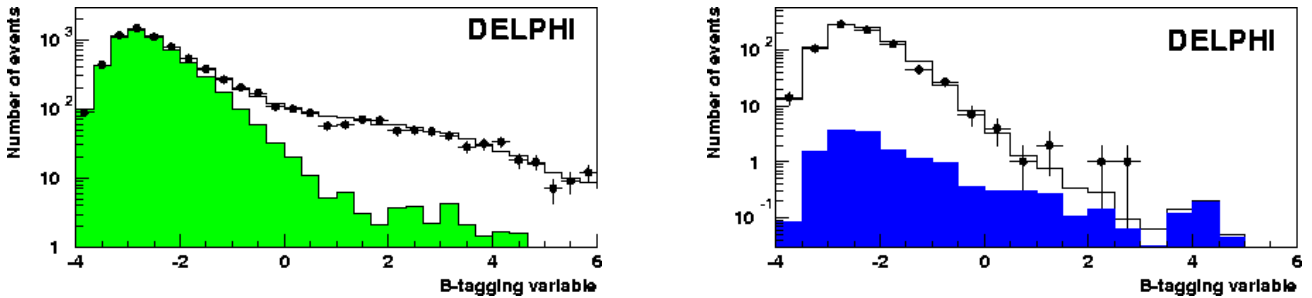

Figure 2. Left: Distributions of the combined b-tag variable, for the year 2000 radiative return $Z\gamma$ data (dots) and simulation (histogram). The expected contribution from udsc-quarks and non $qq\gamma$ background is shown as the dark histogram. Right: Same distribution for semileptonc $W^+W^-$ high energy events in the 2000 data. The shaded histogram is the expected contribution from other processes, and shows the high purity of the selection.

*1.2 Jet pairing and mass reconstruction*

The LEP detectors have a good energy resolution (~20MeV) but even better angular resolution (~2°). Thus a 4C-kinematic fit is imposed, in which energy and mometum are conserved. Effectively the 5$^{th}$ constraint of HZ production is also imposed; D,L,O impose this directly in the fit, whilst A impose it by saying the reconstructed Higgs mass = $m_{12} + m_{34} - m_Z$ where $m_{12}$ and $m_{34}$ are the fitted Z and Higgs boson masses. Slightly different information, including the quality of the b-tag of the jets and the information from the 5C fit, is used by the four experiments to identify the H and Z jets.

*1.3 Analysis procedure*

A basic preselection is applied to select suitable four-jet events. Additional information, such as b-tag, output from the kinematic fit etc along with specific anti QCD, WW and ZZ variables are then combined into a likelihood or neural net to produce a final 'discriminant' variable. The neural net output for ALEPH is shown in figure 3 at the preselection level – good agreement between data and Monte Carlo is seen in the background dominated region.

By combining the reconstructed Higgs mass for a given event with its value in this final discriminant variable we can see how signal or background like it is for a given Higgs boson mass hypothesis. As described in these proceedings[6,7] this is typically expressed as the log of the

likelihood ratio(Q) $\qquad -2\ln(Q) = 2s_{tot} - 2\sum_{i=1}^{N} n_i \ln(1+s_i/b_i)$

where $n_i$ is the number of observed events and $s_i$ and $b_i$ are the expected number of signal and background in a given bin respectively and $s_{tot}$ is the total expected signal. Thus the final value of Q is simply the sum of the individual candidate weights.

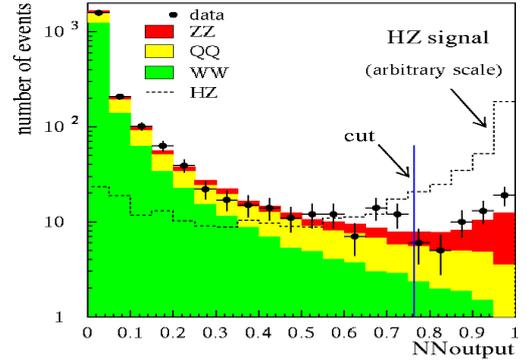

Figure 3 Distribution of the neural network output used to select four-jet HZ candidates for data (dots with error bars), simulated background and simulated signal ($m_H$=114GeV/$c^2$) with an arbitrary normalization, after preselection criteria have been applied.

## 2 Results

Those candidates with a weight greater than 0.3 at 115GeV are shown in table 1. Two of the three most significant candidates are in the four-jet channel, but as discussed this is unsurprising. Indeed the breakdown by channel of the candidates in table 1 is close to that expected from the theoretical branching ratios. For candidate details see refs [1,2,3,4].

| (s/b)$_{115}$ | $M_{rec}$(GeV/$c^2$) | Channel | Expt. |
|---|---|---|---|
| 4.7 | 114 | Hqq | ALEPH |
| 2.3 | 112 | Hqq | ALEPH |
| 1.6 | 114 | Hνν | L3 |
| 0.90 | 110 | Hqq | ALEPH |
| 0.70 | 111 | Hqq | OPAL* |
| 0.60 | 118 | Hee | ALEPH |
| 0.50 | 115 | Hττ | ALEPH |
| 0.50 | 115 | Hqq | ALEPH |
| 0.49 | 113 | Hqq | OPAL |
| 0.49 | 114 | Hνν | L3 |
| 0.45 | 97 | Hqq | DELPHI |
| 0.42 | 115 | Hqq | L3 |
| 0.40 | 114 | Hqq | DELPHI |
| 0.32 | 104 | Hνν | OPAL |

Table 1 Candidates with (s/b)$_{115}$ > 0.3.

The negative log-likelihood ratio as a function of Higgs mass for the four-jet channel is shown in figure 4a. The coloured bands around the background only hypothesis correspond to 1 and 2σ respectively. From this we can see that with this channel there is significant separation (~2σ) of the background only and signal + background hypotheses, even for such a Higgs boson mass of ~115GeV. The data clearly favour the signal+ background hypothesis. The compatibility of the data with the background only hypothesis can be determined and this is shown in figure 4b. The excess with respect to the expected background around 115GeV corresponds to ~2.5σ.

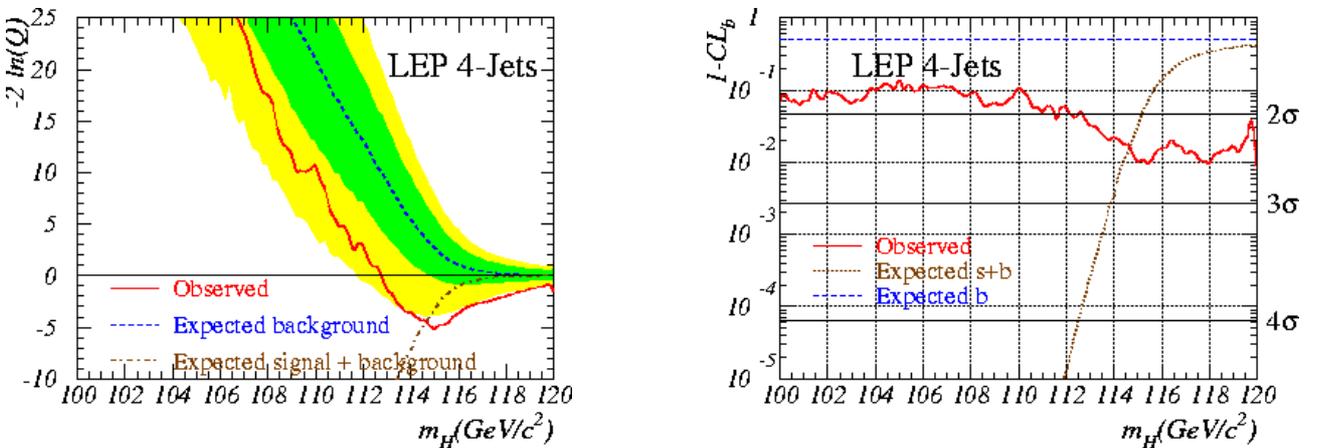

Figure 4 Left plot: Log-likelihood estimator as function of Higgs mass for the observation (solid), background-only (dashed) and signal (dotted dashed) expectations. Right plot: Observed (solid) and expected curves for background (dashed) and signal (dot-dashed) expectations as a function of the Higgs mass.

*3.1 Supporting Evidence*

The four-jet excess is mainly driven by ALEPH. Here the SM Higgs search is carried out in two alternative analysis streams, which are both run blind on the data: a neural network-based stream (NN) and a cut-based stream (cut). The NN specific analyses use neural networks in their selection and two-dimensional discriminants in the calculation of the confidence levels. The cut specific analyses use only the reconstructed Higgs boson mass as a discriminant. Both streams share the same leptonic final state analyses. Although the NN stream is more performant (0.5 GeV/$c^2$ in sensitivity to $m_H$), the cut stream provides an important complementary approach. Both streams see an excess, corresponding to 3.0 σ and 3.1σ in the NN and cut streams respectively. In the NN stream it is driven by the three events listed at the top of Table1. Whilst these events are selected in the cut stream, they do not dominate alone the confidence level as only the mass enters as a discriminant. In light of the excess, additional systematics beyond those usually carried out have been made, including confirming that there is no threshold or b-tag bias in the mass reconstruction[1]. Though work is still in progress to complete the full systematics, the largest expected impact is 0.2σ.

## 4 Conclusion

The four-jet channel is as powerful as the other three channels combined when searching for a SM Higgs boson of mass ~115GeV; but it should not be taken alone, as indeed nor should any single experiment. An excess of events with respect to the expected background, corresponding to ~2.5 σ has been observed for a Higgs boson of mass ~115GeV. It is stable. We look forward to the final experimental publications and the new LEP combination planned for this summer. In the four-jet channel this will use the OPAL event marked by the asterik in table 1 for the first time.

## 5 Acknowledgements

We thank the accelerator divisions for the very successful running of LEP at high energies. Without their extraordinary achievements the excitement of the 2000 data taking would not have been possible.